\documentclass[twocolumn,aps,prl,superscriptaddress]{revtex4-1}
\usepackage{amssymb} 
\usepackage{amsmath,bm} 
\usepackage{graphicx} 
\usepackage[normalem]{ulem} 
\usepackage{multirow} 
\usepackage[colorlinks,linkcolor=blue,urlcolor=blue,citecolor=blue]{hyperref} 
\usepackage{lipsum} 
\usepackage[usenames, dvipsnames]{xcolor} 
\usepackage{tensor} 
\usepackage{isotope} 
\usepackage{amsmath} 
\allowdisplaybreaks[4] 

\renewcommand{\sout}{\bgroup \color{red} \ULdepth=-.5ex \ULset}
   
\begin{document}
\title{Jet-Induced Enhancement of Deuteron Production in $pp$ and $p$-Pb Collisions at the LHC}

\author{Yi-Heng Feng}
\affiliation{Department of Physics and Center for Field Theory and Particle Physics, Fudan University, Shanghai 200438, China}
\affiliation{Key Laboratory of Nuclear Physics and Ion-beam Application (MOE), Institute of Modern Physics, Fudan University, Shanghai 200433, China}

\author{Che Ming Ko} 
\email{ko@comp.tamu.edu} 
\affiliation{Cyclotron Institute and Department of Physics and Astronomy, Texas A\&M University, College Station, Texas 77843, USA}

\author{Yu-Gang Ma}
\email{mayugang@fudan.edu.cn}
\affiliation{Key Laboratory of Nuclear Physics and Ion-beam Application (MOE), Institute of Modern Physics, Fudan University, Shanghai 200433, China}
\affiliation{Shanghai Research Center for Theoretical Nuclear Physics, NSFC and Fudan University, Shanghai 200438, China}

\author{Kai-Jia Sun}
\email{kjsun@fudan.edu.cn}
\affiliation{Key Laboratory of Nuclear Physics and Ion-beam Application (MOE), Institute of Modern Physics, Fudan University, Shanghai 200433, China}
\affiliation{Shanghai Research Center for Theoretical Nuclear Physics, NSFC and Fudan University, Shanghai 200438, China}

\author{Xin-Nian Wang}
\email{xnwang@mail.ccnu.edu.cn}
\affiliation{Key Laboratory of Quark and Lepton Physics (MOE) \& Institute of Particle Physics, Central China Normal University, Wuhan 430079, China}

\author{Zhong Yang}
\email{yangzhong1994@mails.ccnu.edu.cn}
\affiliation{Key Laboratory of Quark and Lepton Physics (MOE) \& Institute of Particle Physics, Central China Normal University, Wuhan 430079, China} 

\author{Song Zhang}
\email{song\_zhang@fudan.edu.cn}
\affiliation{Key Laboratory of Nuclear Physics and Ion-beam Application (MOE), Institute of Modern Physics, Fudan University, Shanghai 200433, China}
\affiliation{Shanghai Research Center for Theoretical Nuclear Physics, NSFC and Fudan University, Shanghai 200438, China}

\date{\today}

\begin{abstract}
Jet-associated deuteron production in  $pp$ collisions at $\sqrt{s}=13$ TeV and $p$-Pb collisions at $\sqrt{s_{NN}}=5.02$ TeV  is studied in the coalescence model by using the phase-space information of proton and neutron pairs from a multiphase transport (AMPT)  model at the kinetic freezeout. In the low transverse momentum ($p_T$) region $p_T/A < 1.5$ GeV/$c$, where $A$ is the mass number of a nucleus, the in-jet coalescence factor $B_2^\text{In-jet}$ for deuteron production, given by the ratio of the in-jet deuteron number to the square of the in-jet proton number, is found to be  larger than the coalescence factor $B_2$  in the medium perpendicular to the jet by a factor of about 10 in $pp$ collisions and of 25 in $p-$Pb collisions, which are consistent with the ALICE measurements at the LHC.  Such large low-momentum enhancements mainly come from coalescence of  nucleons inside the jet with the medium  nucleons. Coalescence of nucleons inside the jet  dominates deuteron production only at the higher $p_T$ region of $p_T/A\gtrsim 4$ GeV/$c$, where  both the yield ratio $d/p$ of deuteron to proton numbers and the $B_2$ are also significantly larger in the jet direction than in the direction perpendicular to the jet due to the strong collinear correlation among particles produced from jet fragmentation. Studying jet-associated deuteron production in relativistic nuclear collisions thus opens up a new window to probe the phase-space structure of nucleons inside jets.
\end{abstract} 

\maketitle

\emph{Introduction.}{\bf ---}
Light nuclei, such as deuteron ($d$), helium ($^3$He and $^4$He), and their anti-particles, are observed in high-energy nuclear collisions at the Relativistic Heavy Ion Collider (RHIC) and the Large Hadron Collider (LHC)~\cite{ALICE:2020foi,STAR:2010gyg,STAR:2011eej,ALICE:2015rey,STAR:2019wjm,ALICE:2015wav,STAR:2009tro}. These bound states have small binding energies and appreciable sizes,  compared with the temperature and size of the produced fireball in high-energy nuclear collisions. The study of light nuclei production is important due to its relevance  to  the indirect dark matter detection~\cite{Hailey:2009fpa,Kounine:2012ega,Donato:2008yx,Korsmeier:2017xzj,AMS:2016oqu,ALICE:2022zuz} and  the search for the possible critical point in the phase diagram of strongly interacting matter ~\cite{Luo:2017faz,Sun:2017xrx,Shuryak:2018lgd,Yu:2018kvh,Sun:2020zxy,Sun:2020pjz,Sun:2020bbn,Sun:2018jhg,STAR:2022hbp,MaYG,KoCM,ZhangY,SunKJ,ZhuLL}.

Recently, the ALICE Collaboration at the LHC has measured the jet-associated deuterons in $pp$ collisions at $\sqrt{s}=13$ TeV and found the deuteron formation probability inside the jet cone to be an order of magnitude larger than that in the transverse region perpendicular to the jet direction~\cite{ALICE:2020hjy,ALICE:2022ugx}. This large enhancement of deuteron production has been attributed to the strong correlation among particles produced from jet fragmentation, i.e., particles close in space also have similar momentum~\cite{Field:1977fa,Andersson:1983ia}, which then results in a large probability for the formation of a deuteron if it is produced from the coalescence of proton and neutron pairs in the jet. The nucleosynthesis in jets thus provides a connection between hard and soft processes in high-energy nuclear collisions.

The production mechanism of light nuclei in relativistic nuclear collisions is currently under intense debate \cite{Oh:2009gx,Feckova:2016kjx,Mrowczynski:2016xqm,Bazak:2018hgl,Oliinychenko:2018ugs,Bellini:2018epz,Bugaev:2018klr}. According to the statistical hadronization model (SHM) \cite{Vovchenko:2019kes,Andronic:2010qu,Andronic:2016nof}, light nuclei are produced at a common chemical freeze-out temperature after hadronization of the quark-gluon plasma (QGP). In the nucleon coalescence model~\cite{Scheibl:1998tk,Sato:1981ez,ZhangYX,WangTT,Sun:2018mqq}, light nuclei are formed at the kinetic freeze-out stage, when the temperature and density of the hadronic matter are much lower. There is also the recently developed kinetic model that takes into account the dissociation and regeneration of light nuclei through pion-catalyzed multi-nucleon reactions during the expansion of the hadronic matter~\cite{Sun:2022xjr,Wang:2023gta,Oh:2007vf,Oliinychenko:2018ugs,Danielewicz:1991dh}. Studying light nuclei production in jets may provide valuable information on their production mechanism.

\begin{figure}[!t]
 \centering
\includegraphics[width=0.34\textwidth,trim = {-30 0 0 0},clip]{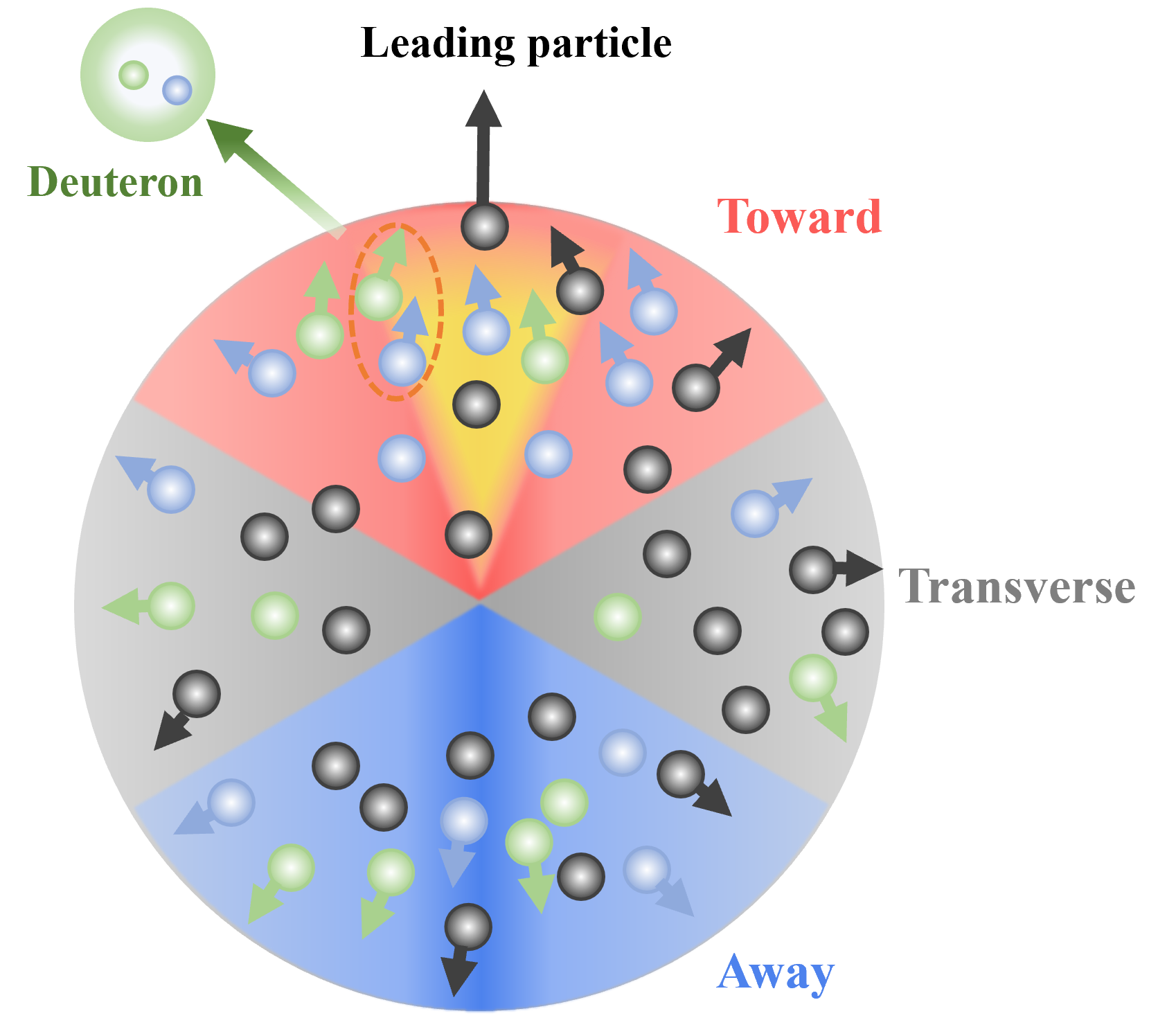}
 \caption{Jet-associated deuteron production in different transverse regions relative to the leading particle:  $|\Delta \phi|<60^\circ$ (Toward),  $|\Delta \phi|>120^\circ$ (Away), and  $60^\circ<|\Delta \phi|<120^\circ$ (Transverse), with $\Delta \phi$ being the azimuthal angle of the emitted particle relative to that of the leading particle. }
 \label{pic:jets}
\end{figure}

The ALICE measurements of the jet-associated deuterons are at low $p_T$ region ($p_T/A\le 1.5$ GeV/$c$) where the fraction of protons from the jet is only about $10\%$~\cite{ALICE:2023yuk}. Although the coalescence factor $B_2$ in this transverse momentum region has similar values for deuteron production in the Toward region with azimuthal angles $|\Delta\phi|<60^\circ$ around the jet direction and the Transverse region of $60^\circ<|\Delta \phi|<120^\circ$, shown in Fig.~\ref{pic:jets}, the jet effect on deuteron production is, however, amplified by considering the in-jet coalescence factor for deuteron production, 
\begin{eqnarray}
B_2^{\text{In-jet}} = \left(\frac{1}{2{\pi}p^{d}_T} \frac{\mathrm{d}^2N_{d}^{\text{In-jet}}}{\mathrm{d}y\mathrm{d}p^{d}_T}\right)\bigg{/}\left(\frac{1}{2{\pi}p^{p}_T} \frac{\mathrm{d}^2N_{p}^{\text{In-jet}}}{\mathrm{d}y\mathrm{d}p^{p}_T}\right)^2, 
\end{eqnarray}
where the in-jet proton ($N_{p}^{\text{In-jet}}$) and deuteron ($N_{d}^{\text{In-jet}}$) numbers are defined as the difference between their respective numbers in the Toward and Transverse regions, i.e., $N_{p,d}^{\text{In-jet}}=N_{p,d}^{\rm Toward}-N_{p,d}^{\rm Transverse}$.

In the present study, we study the jet effect on deuteron production in $pp$ collisions at $\sqrt{s} = 13$ TeV and in $p$-Pb collisions at $\sqrt{s_{NN}} = 5.02$ TeV in both the low $p_T$ region and the higher $p_T$ region ($p_T/A>2$ GeV/$c$) to see its dependence on the deuteron momentum and the size of collision system. We find that both deuteron production from the jet-medium coalescence of nucleons in the jet with nucleons in the medium as well as from jet-jet coalescence of nucleons only in the jet exceeds that from the medium-medium coalescence for $p_T/A\gtrsim 3$ GeV/$c$. The jet-jet coalescence dominates at $p_T/A\gtrsim 4$ GeV/$c$, where both the yield ratio $N_d/N_p$ or $d/p$ of deuteron to proton numbers and the coalescence factor $B_2$  are significantly larger in the Toward region than in the Transverse region as a result of the increasingly narrower neutron-proton pair angular distribution inside the jet cone. Also found is the even larger in-jet coalescence factor $B_2^{\rm In-jet}$ in $p$-Pb collisions than in $pp$ collisions.

\emph{Methods.}{\bf ---}In the present study, we use the AMPT model \cite{Lin:2004en,LinZW} to simulate the evolution of a collision system at relativistic energies and obtain the final phase-space distribution of kinetically freeze-out nucleons. This model has been widely adopted for studying light nuclei production and jet observables, such as the jet shape \cite{Luo:2021hoo}, jet quenching effects \cite{Ma:2013pha, Luo:2021voy, Ma:2010dv, Duan:2023gmp}, deuteron production and deuteron elliptic flow \cite{Oh:2009gx,Shao:2022eyd}, light (hyper)nuclei production and correlation \cite{ZhangS0,ChengYL,WangTT} as well as nuclear structure effects \cite{ZhangS,LiYA,WangYZ} and so on. The AMPT model contains four main stages: initial condition, parton cascade, hadronization, and hadronic decays and rescatterings. The initial condition of the AMPT model is obtained from the HIJING model \cite{Wang:1991hta,Gyulassy:1994ew} by converting hadrons produced from minijets and soft string excitations to their valence quarks and antiquarks. The evolution of these partons is then described by Zhang's Parton Cascade (ZPC) model \cite{Zhang:1997ej}. For the hadronization of this quark matter after partons stop scattering or its local energy density decreases to a critical value, a quark coalescence model is used \cite{Lin:2001zk}, which is followed by a relativistic transport (ART) model \cite{Li:1995pra} to describe the hadronic scatterings and decays of resonances. With the phase-space information of nucleons at kinetic freeze-out, we then study light nuclei production using the nucleon coalescence model.

In the nucleon coalescence model for light nuclei production~\cite{Scheibl:1998tk,Sun:2018mqq}, the formation probability of a deuteron from a proton and neutron pair is given by the Wigner function \cite{Wigner:1932eb} of the deuteron internal wave function, which we take as
\begin{equation}
W_d(\bm{x}, \bm{p}) = 8g_de^{-\frac{\bm{x}^2}{\sigma^2}-\bm{p}^{2}\sigma^2},
\end{equation}
where $g_{d}=3/4$ is the statistical factor for spin 1/2 proton and neutron to form a spin 1 deuteron, and  ${\bf x}=({\bf x}_1-{\bf x}
_2)/\sqrt{2}$ and ${\bf p}=({\bf p}_1-{\bf p}_2)/\sqrt{2}$ are the relative coordinate and momentum, respectively. Here, ${\bf x}_1$ 
and ${\bf x}_2$ as well as ${\bf p}_1$ and ${\bf p}_2$ are the spatial coordinates and momenta of the two nucleons in their rest frame, with ${\bf x}_1$ and ${\bf x}_2$ taken at equal time by propagating the nucleon with an earlier freeze-out time to the time of the later freeze-out nucleon. For the size parameter $\sigma_{d}$ in the Wigner function, it is related to the deuteron root-mean-squared radius  by $\sigma_{d}  = \sqrt{4/3}~r_{d}\approx 2.26$ fm~\cite{Sun:2017ooe,Ropke:2008qk}.
 
\emph{Jet effects on deuteron production.}{\bf ---} As in experimental measurements~\cite{ALICE:2023yuk,ALICE:2020hjy}, we select events with a leading particle in the pseudorapidity $|\eta| < 0.9$ that has $p_T$ greater than $5$ GeV/$c$. With the direction of the leading particle taken to approximate the direction of the jet, the transverse plane is then divided into three azimuthal regions shown in Fig.~\ref{pic:jets}, which  include the Away region of azimuthal angles $|\Delta \phi|>120^\circ$ besides the Toward and Transverse regions defined in the previous Section.
 
\begin{figure}[!t]
\centering
\includegraphics[width=8.55cm]{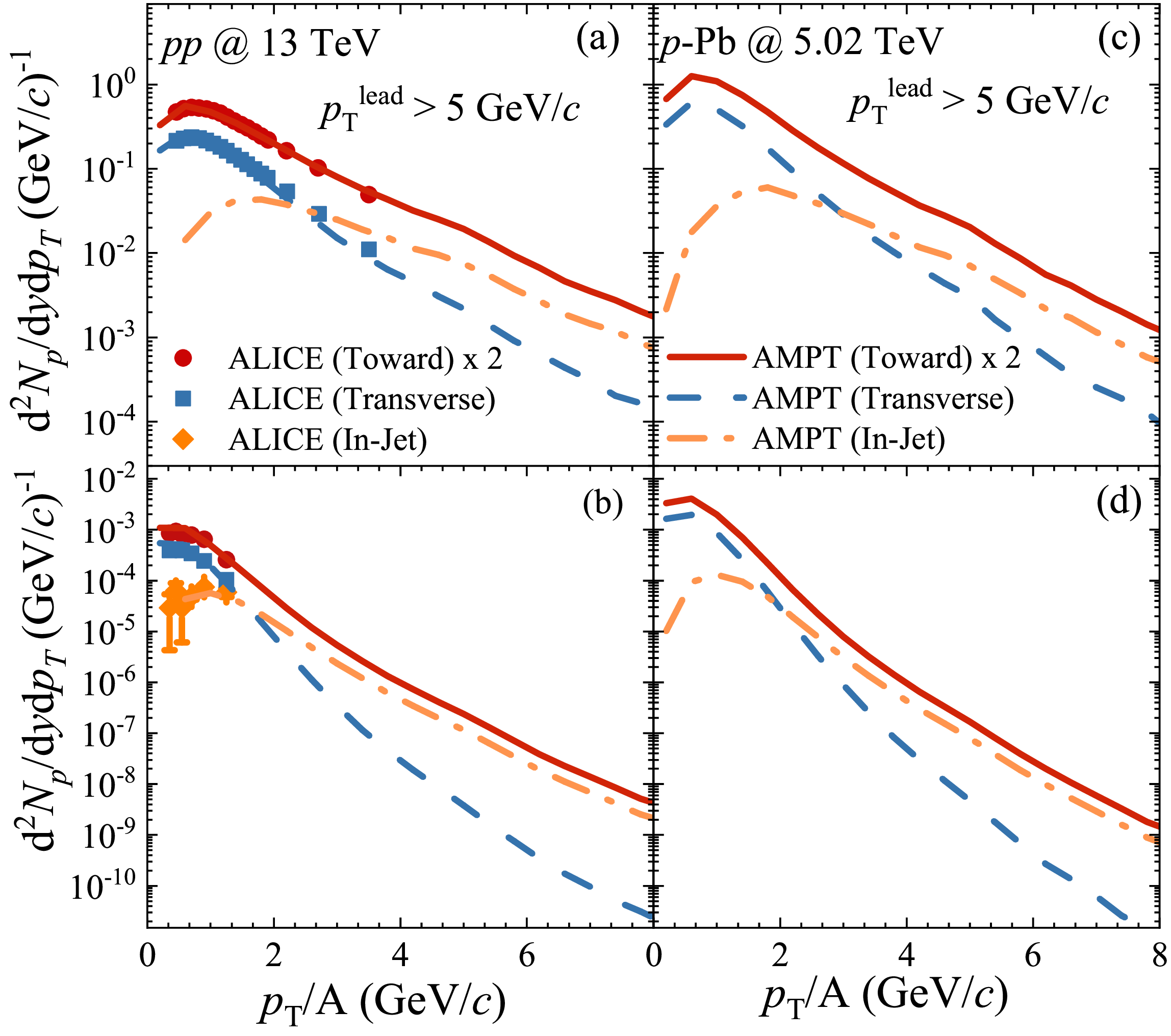}
\caption{Transverse momentum spectrum in different azimuthal regions from the AMPT model and comparison with experimental data \cite{ALICE:2023yuk,ALICE:2022ugx}. (a) Proton in $pp$ collisions, (b) deuteron in $pp$ collisions, (c)proton in $p$-Pb collisions, and (d)deuteron in $p$-Pb collisions.}
\label{pic:spectra}
\end{figure}

Figure~\ref{pic:spectra} (a) presents the proton transverse momentum spectrum in various azimuthal regions obtained from the AMPT model for $pp$ collisions,  while Fig.~\ref{pic:spectra} (b) displays the  same for the deuteron transverse momentum spectrum obtained from the coalescence model using kinetically freeze-out nucleons from AMPT. The AMPT model is seen to reproduce the experimental data in the Toward region and slightly underestimate the data in the Transverse region. Also depicted in Fig.~\ref{pic:spectra} (c) and (d) are the proton and deuteron transverse momentum spectra in the Toward and Transverse regions from $p-$Pb collisions, which are seen to be larger than corresponding ones in $pp$ collisions. However, the in-jet proton and deuteron transverse momentum spectra, defined as the difference between their respective numbers in the Toward and Transverse regions, from the two collision systems are  very close.

The jet effects on deuteron production can be more clearly seen from the ratio of deuteron and proton transverse momentum spectra as well as the deuteron coalescence factor,i.e., 
\begin{eqnarray}
d/p &=& \left(\frac{1}{2{\pi}p^\mathrm{d}_T} \frac{d^2N_\mathrm{d}}{dydp^\mathrm{d}_T}\right)\bigg{/}\left(\frac{1}{2{\pi}p^\mathrm{p}_T} \frac{d^2N_\mathrm{p}}{dydp^\mathrm{p}_T}\right),\\
B_2 &=& \left(\frac{1}{2{\pi}p^\mathrm{d}_T} \frac{d^2N_\mathrm{d}}{dydp^\mathrm{d}_T}\right)\bigg{/}\left(\frac{1}{2{\pi}p^\mathrm{p}_T} \frac{d^2N_\mathrm{p}}{dydp^\mathrm{p}_T}\right)^2,
\end{eqnarray}
where $p^d_T = 2p^p_T$. 

\begin{figure}[!t]
\centering
\includegraphics[width=8.5cm]{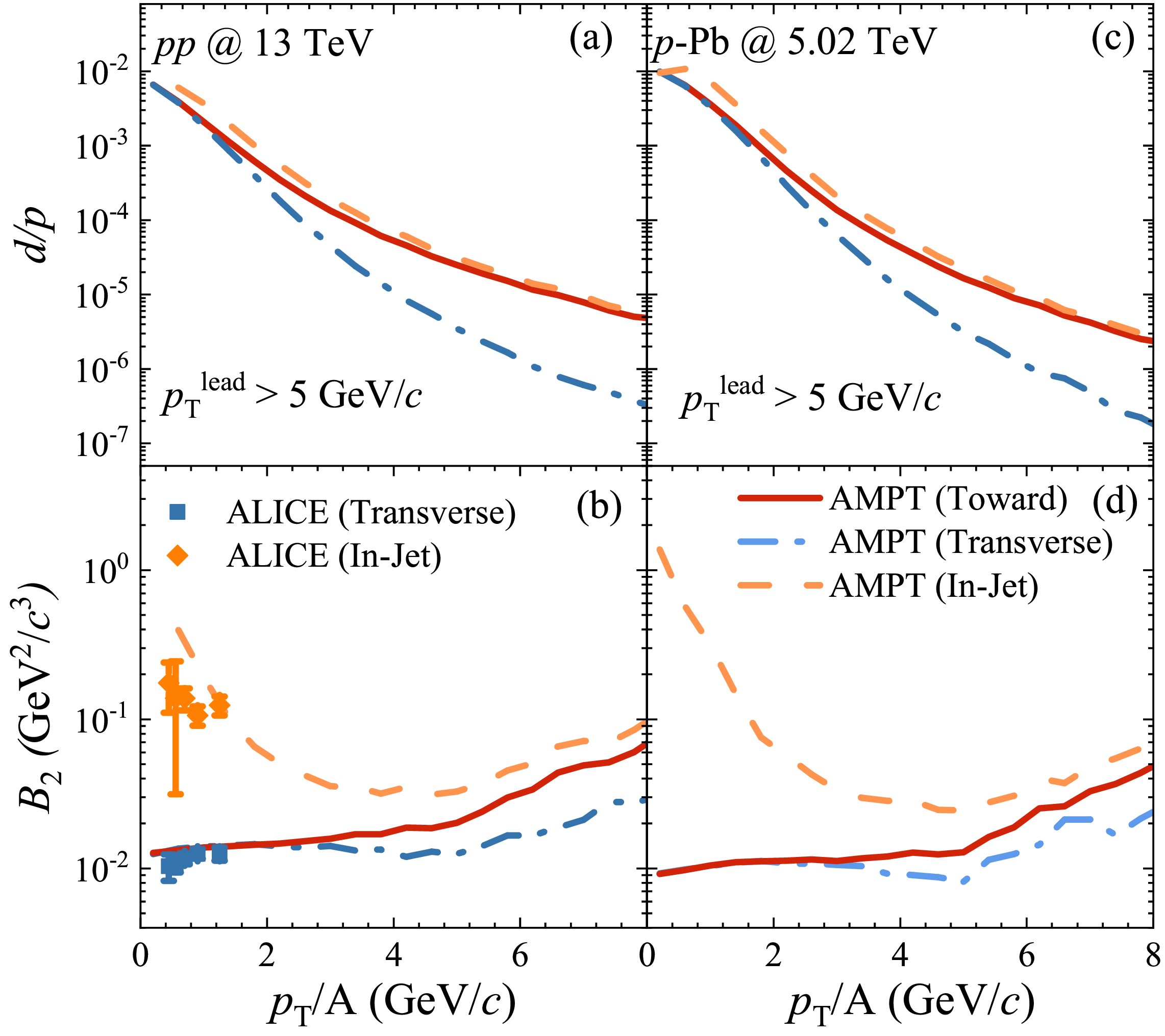}
  \caption{ $d/p$ ratio and $B_2$ in different azimuthal regions from the AMPT model and comparison with experimental data \cite{ALICE:2022ugx}. (a) $d/p$ ratio in $pp$ collisions, (b) $B_2$ in $pp$ collisions, (c) $d/p$ ratio in $p$-Pb collisions, and (d) $B_2$ in $p$-Pb collisions. }
  \label{pic:B2}
\end{figure}

\begin{figure}[!b]
\centering
\includegraphics[width=8.5cm]{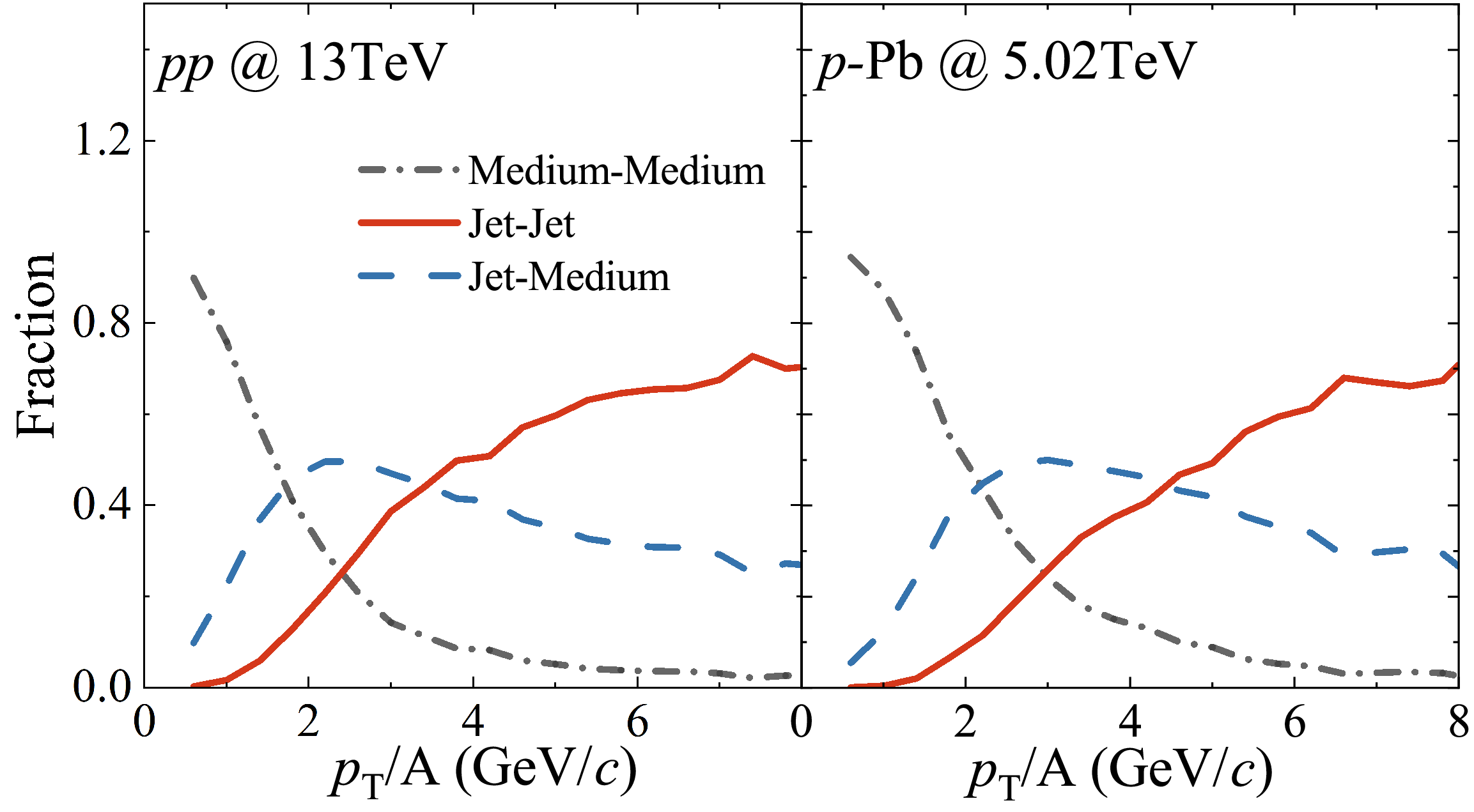}
  \caption{Fractions of deuterons produced by Jet-Jet nucleon coalescence (solid line), Medium-Medium nucleon coalescence (dash-dotted line), and Jet-Medium nucleon coalescence (dashed line) as a function of transverse momentum.}
  \label{pic:fraction}
\end{figure}

\begin{figure*}[!t]
\centering
\includegraphics[width=16cm]{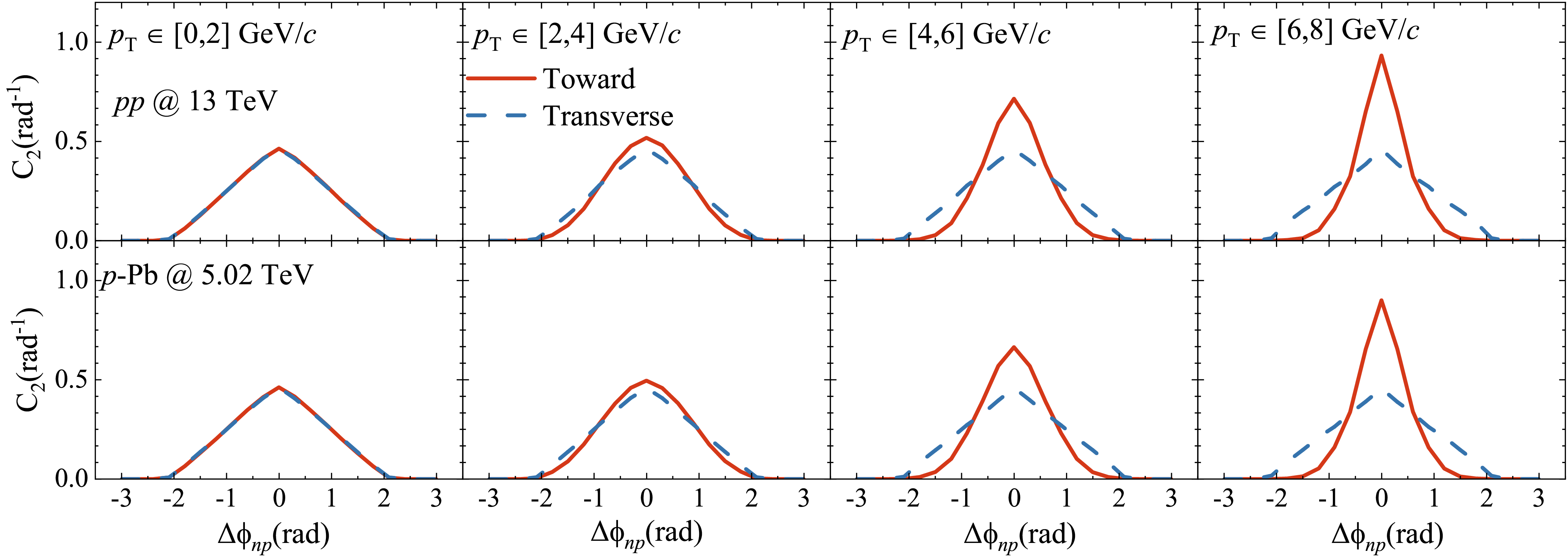}
  \caption{ Distribution of the difference in the azimuthal angles of neutron and proton pairs in the Toward and Transverse regions for $pp$ collisions at $\sqrt{s}=13$ TeV (upper panels) and $p-$Pb collisions at $\sqrt{s_{NN}}=5.02$ TeV (lower panels). The upper (lower) four panels correspond to four different regions of the transverse momentum of neutron and proton pairs.}
  \label{pic:AngleDist}
\end{figure*}

Figure~\ref{pic:B2} (a) and (c) show the yield ratio $d/p$ in different azimuthal regions for $pp$ and $p-$Pb collisions, respectively. It is seen that its value decreases with increasing $p_T$ in both  Toward and  Transverse regions, with a larger value in the former than in the latter, and their difference increases gradually with increasing $p_T$, resulting in its in-jet value close to that in the Toward region at $p_T>2$ GeV/$c$. The deuteron coalescence factor $B_2$, which also increases with increasing $p_T$ in both Toward and Transverse regions, exhibits a similar behavior as shown in Fig.~\ref{pic:B2} (b) and (d).  At low momenta of $p_T/A < 1.5$ GeV/$c$, the \( B_2 \) values in the Toward and Transverse regions are almost identical and this is in agreement with a recent study~\cite{Bailung:2024sca}, which demonstrates that the increase in \( B_2 \) of "jetty" deuterons is negligible. However, the in-jet deuteron coalescence factor $B^{\rm In-jet}_2$ is larger than the $B^{\rm Transverse}_2$ in the Transverse region by a factor of about 10 in $pp$ collisions, which is similar to that measured in the experiment by the ALICE Collaboration~\cite{ALICE:2022ugx}, and about 25 at $p_T/A = 1$ GeV/$c$ in $p-$Pb collisions. This order of magnitude larger $B^{\rm In-jet}_2$ than the $B^{\rm Transverse}_2$ in the Transverse region at low $p_T$ can be understood from the number of deuterons produced in the Toward region. Since nucleons in this region ($N^{\rm Toward}_{p,n}$) consist of nucleons from both underlying events or the medium ($N^{\rm Medium}_{p,n}$) and the jet ($N^{\rm In-jet}_{p,n}$), i.e., $N^{\rm Toward}_{p,n}=N^{\rm Medium}_{p,n}+N^{\rm In-jet}_{p,n}$, deuterons in the Toward region can thus be produced from Jet-Jet nucleon coalescence, Medium-Medium nucleon coalescence, and Jet-Medium nucleon coalescence.  

As $N_n \approx N_p$, the fractions of these three contributions to deuteron production in the Toward region are
\begin{equation}
\begin{aligned}
    &{\rm Fraction_{Jet}} = \frac{(N^{\rm In-jet}_p)^2}{(N^{\rm Toward}_p)^2}, \\ 
    &{\rm Fraction_{Medium}} =  \frac{(N^{\rm Medium}_p)^2}{(N^{\rm Toward}_p)^2}, \\   
    &{\rm Fraction_{Jet-Medium}} = 2\frac{N^{\rm In-jet}_p \times N^{\rm Medium}_p}{(N^{\rm Toward}_p)^2}.
\end{aligned}
\end{equation}
Figure~\ref{pic:fraction} shows the fractions of contribution from the three terms.  At $p_T/A < 1.5$ Gev/$c$, the contribution of Jet-Jet nucleon coalescence is much smaller than that of Jet-Medium nucleon coalescence, which is further much smaller than that of Medium-Medium nucleon coalescence.  The contribution from Jet-Jet nucleon coalescence becomes, however, dominant at $p_T/A> 4$ GeV/$c$. 

The in-jet coalescence factor $B_2^{\rm In-jet}$ at $p_T/A < 1.5$ GeV/$c$ is thus approximately given by 
\begin{eqnarray}\label{coalfac}
B^{\rm In-jet}_2&\approx &\frac{N^{\rm Jet-Medium}_d}{(N^{\rm In-jet}_p)^2}=\frac{2N^{\rm Medium}_p}{N^{\rm In-jet}_p}\times B^{\rm Toward}_2\notag\\
&\approx & \frac{2 N^{\rm Transverse}_p}{N^{\rm In-jet}_p}\times B^{\rm Toward}_2,
\end{eqnarray}
where we have used  $N^{\rm Medium}_p=N^{\rm Transverse}_p$. Because of $N^{\rm Transverse}_p\gg N^{\rm In-jet}_p$ at $p_T/A < 1.5$ GeV/$c$, $B^{\rm In-jet}_2$ is much larger than $B^{\rm Toward}_2$ and $B^{\rm Transverse}_2$ in low $p_T$ region.   With a larger $N^{\rm Transverse}_p/N_p^{\rm In-jet}$ in $p$-Pb collisions than in $pp$ collisions, an even larger $B^{\rm In-jet}_2$ is expected in this collision as shown in Fig.~\ref{pic:B2}.

Although the $B_2^{\rm In-jet}$ has a very large value at low $p_T$, it does not provide any phase-space information of the nucleons in a jet. To achieve this, it is better to compare directly deuteron production in the Toward and Transverse regions as Fig.~\ref{pic:B2} shows that the difference in the $B_2$ between the Toward and Transverse regions increases with increasing $p_T$ and becomes appreciable at $p_T\ge 3$ GeV/$c$.  The larger value of $B_2$ in the Toward region than the Transverse region at $p_T>3$ GeV/$c$ is because particles are more collimated in the Toward region. This can be understood by considering  the   distribution of     the azimuthal angle difference $\Delta\phi_{np}=\phi_n-\phi_p$ between the proton and neutron azimuthal angles $\phi_p$ and $\phi_n$  in momentum space, which we define as
\begin{eqnarray}
C_2(\Delta \phi_{np}) = \frac{\text{d}N_{\text{pair}}}{\text{d}\Delta \phi_{np}}. 
\end{eqnarray}
To form a deuteron requires the momentum difference between the neutron and proton to be about  $r_d^{-1}\approx 100$ MeV. As a result, for a  deuteron  of sufficiently large $p_T$, the value of the  $\Delta\phi_{np}$  between its  proton and neutron  approaches zero. In this case, one expects the coalescence factor to be proportional to the value of $C_2(\Delta\phi_{np})$ at $\Delta\phi_{np} = 0$, i.e., $B_2\propto C_2(\Delta\phi_{np} = 0)$.

Figure~\ref{pic:AngleDist} displays the distribution of the azimuthal angle difference between kinetically freeze-out neutron and proton pairs in the AMPT as a function of $p_T/A$ in both $pp$ collisions at $\sqrt{s}=13$ TeV (upper panels) and $p-$Pb collisions at $\sqrt{s_{NN}}=5.02$ TeV (lower panels).  At $p_T/A<2$ GeV/$c$, the $\Delta\phi_{pn}$  distribution has  a triangular shape because of the restricted angular  range of $2/3\pi$, and it is almost identical in the Toward and Transverse regions. Although the distribution $C_2(\Delta\phi_{np})$ at $p_T/A\ge 2$ GeV/$c$ remains unchanged in the Transverse region, it becomes increasingly narrower in the Toward region as $p_T/A$ increases, suggesting that proton and neutron pairs in the Toward region have a larger coalescence probability when their total momentum is large.  At $p_T/A\gtrsim 4$  GeV/$c$, the $\Delta\phi_{pn}$  distribution in the Toward region becomes significantly sharper,   indicating more collimated nucleons  along the jet direction~\cite{Mrowczynski:2023hbn}   and Jet-Jet coalescence dominated  deuteron production in this region.

\begin{figure}[!t]
\centering
\includegraphics[width=8.5cm]{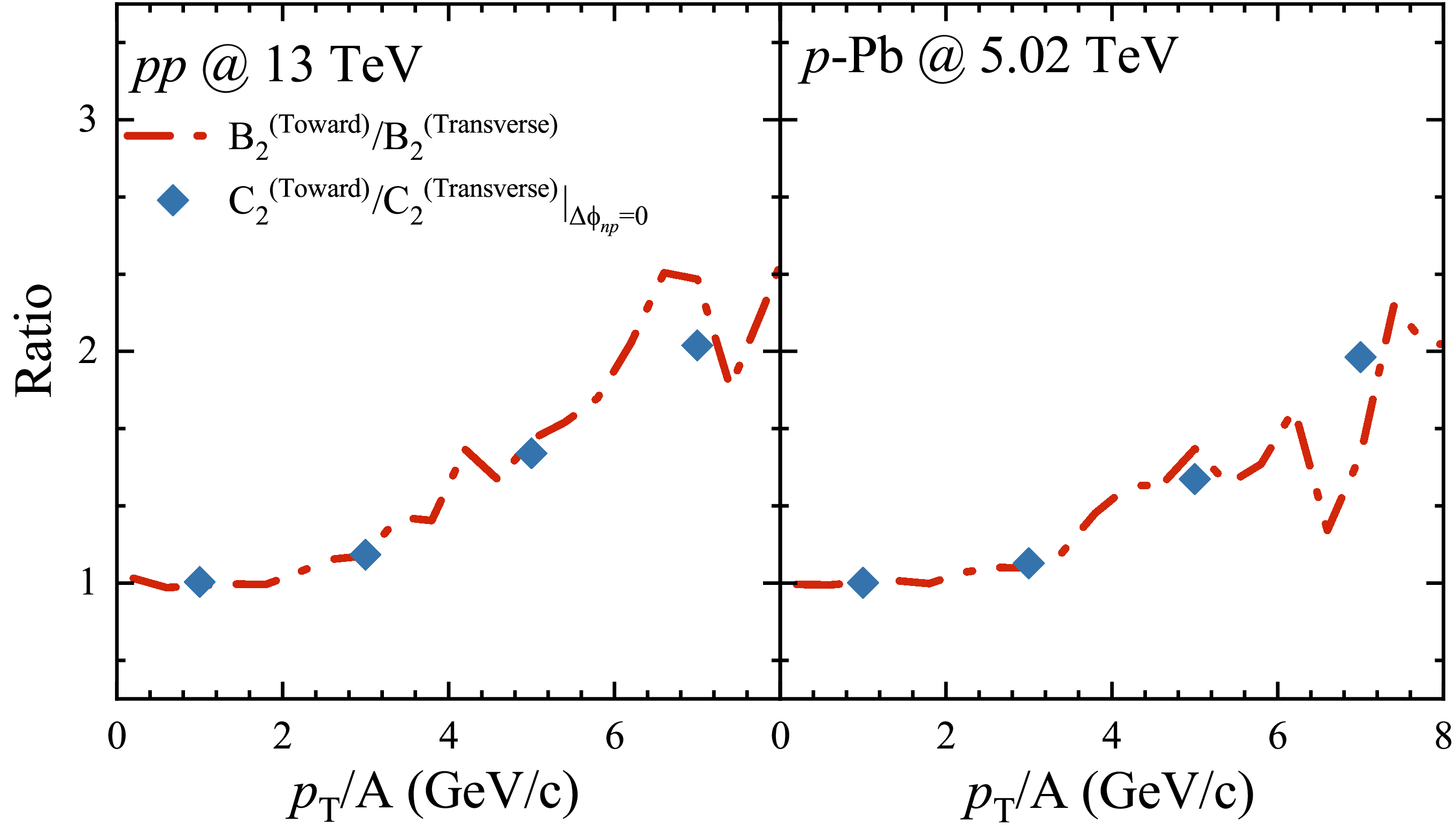}
  \caption{Ratios of coalescence factor $B_2$ and angular distribution function $C_2$ in the Toward region to the respective ones in the Transverse region as a function of $p_T/A$.  Shown in the upper (lower) panel are results for $pp$ ($p-$Pb) collisions. Dashed lines and solid symbols denote ratios of $B_2$ and $C_2$, respectively.}
  \label{pic:ratio}
\end{figure}

Figure~\ref{pic:ratio} displays the ratio of $B_2$ in the Toward region and the Transverse region as well as that of $C_2$ as a function of $p_T/A$. Both ratios increase with increasing $p_T/A$, and they become almost identical in both $pp$ collisions at $\sqrt{s}=13$ TeV (upper panels) and $p-$Pb collisions at $\sqrt{s_{NN}}=5.02$ TeV (lower panels). This result suggests that the increase of $B_2$ at large $p_T$ in the Toward region, shown in Fig.~\ref{pic:B2}, indeed comes from the genuine effect of jets, i.e., particles' momenta are strongly correlated inside the jet cone, pointing towards the direction of the leading particles.

\emph{Summary.}{\bf ---}
In the present study, we investigate jet effects on deuteron production in both $pp$ and $p$-Pb collisions at the LHC energies 
using the nucleon coalescence model for light nuclei production with the nucleon phase-space information obtained from the 
AMPT Model. In the low-momentum region ($p_T/A < 1.5$ GeV/$c$), the in-jet deuteron coalescence factor $B_2^{\rm 
In-jet}$ is found to be enhanced by factors of about 10 in $pp$ and 25 in $p-$Pb collisions, which is consistent with recent ALICE 
measurements. However, we find that such large enhancements mainly come from the jet-medium coalescence, 
not the jet-jet coalescence.  Only at the very higher $p_T$ region ($p_T/A\gtrsim 4$ GeV/$c$), deuteron production 
from the jet-jet coalescence dominates, and both the yield ratio $d/p$ of deuteron to proton numbers and the deuteron coalescence factor $B_2$ are significantly larger in the Toward region than in the Transverse region.

We also find that the enhanced deuteron production at $p_T/A\gtrsim 4$ GeV/$c$ inside a jet cone is directly linked to the distinct neutron and proton pair angular distributions in and out of the jet, providing evidence that deuteron production is sensitive to the phase-space structure of nucleons in the jet. These findings suggest that nucleosynthesis in jets may serve as a promising tool to study the medium response to jet quenching in QGP~\cite{Yang:2022nei,Qin:2023}.   Future experimental and theoretical investigations of jet effects on nucleosynthesis in collisions of both small and large systems will be of great interest.

\emph{Acknowledgments.}{\bf ---}
We thank Zhangbu Xu for insightful discussions on the argument presented in Eq. (6),  Zi-Wei Lin, and Rui Wang for helpful discussions, and Chen Zhong for helping with the server. This work was supported in part by the National Key Research and Development Project of China under Grant No. 2022YFA1602303 and No. 2022YFA1604900 and the National Natural Science Foundation of China under Grant No. 12375121, 12147101, 11891070, 11890714, 12322508 and 11935007  as well as the U.S. Department of Energy under Award No. DE-SC0015266. 
%

\end{document}